\documentclass[aps,prl,twocolumn,groupedaddress,preprintnumbers,nofootinbib]{revtex4}

\usepackage[dvips]{graphicx}
\usepackage{color}
\usepackage{amsmath,amssymb,slashed}
\newcommand{\eqqa}{\begin{equation}\begin{aligned}}
\newcommand{\eqqae}{\end{aligned}\end{equation}}

\newif\ifdraft
\drafttrue
\newif\ifpreprint
\preprinttrue
\def\spa#1.#2{\left\langle#1\,#2\right\rangle}
\def\spb#1.#2{\left[#1\,#2\right]}

\newcommand{\eq}{\begin{equation}}
\newcommand{\eqe}{\end{equation}}
\newcommand{\eqa}{\begin{eqnarray}}
\newcommand{\eqae}{\end{eqnarray}}

\newcommand{\bea}{\begin{eqnarray}}
\newcommand{\eea}{\end{eqnarray}}

\newbox\charbox
\newbox\slabox
\def\s#1{{      % Feynman slash
        \setbox\charbox=\hbox{$#1$}
        \setbox\slabox=\hbox{$/$}
        \dimen\charbox=\ht\slabox
        \advance\dimen\charbox by -\dp\slabox
        \advance\dimen\charbox by -\ht\charbox
        \advance\dimen\charbox by \dp\charbox
        \divide\dimen\charbox by 2
        \raise-\dimen\charbox\hbox to \wd\charbox{\hss/\hss}
        \llap{$#1$}
}}

\begin{document}

\preprint{QMUL-PH-18-21}
\preprint{NCTS-TH/1810}
\title{Dualities for Ising networks} 

\author{
Yu-tin Huang,$^{1,2}$, Chia-Kai Kuo,$^1$,  Congkao Wen,$^{3}$}

\affiliation{$^1$ Department of Physics and Astronomy, National Taiwan University, Taipei 10617, Taiwan}
\affiliation{$^2$ Physics Division, National Center for Theoretical Sciences, National Tsing-Hua University,
No.101, Section 2, Kuang-Fu Road, Hsinchu, Taiwan}

\affiliation{$^2$ Centre for Research in String Theory, School of Physics and Astronomy
Queen Mary University of London, Mile End Road, London E1 4NS, United Kingdom}

\begin{abstract}
In this note, we study the equivalence between planar Ising networks and cells in the positive orthogonal Grassmannian. We present a microscopic construction based on amalgamation, which establishes the correspondence for any planar Ising network. The equivalence allows us to introduce two recursive methods for computing correlators of Ising networks. The first based on duality moves, which generate networks belonging to the same cell in the Grassmannian. This leads to fractal lattices where the recursion formulas become the exact RG equations of the effective couplings. For the second, we use amalgamation where each iteration doubles the size of the seed lattice. This leads to an efficient way of computing the correlator where the complexity scales logarithmically with respect to the number of spin sites. 
\end{abstract}

\maketitle

%%%%%%%%%%%%%%%%%%%%%%%%%%%%%%%%%%%%%%%%%%%%%%%%%%%%%
\section{Introduction}  
%%%%%%%%%%%%%%%%%%%%%%%%%%%%%%%%%%%%%%%%%%%%%%%%%%%%%

\vspace{-0.3cm}
Recent years there has been a fascinating interplay between the physics of observables in quantum field theories and geometries in mathematics. Consistency conditions of the observables, arising from fundamental principles of unitarity, locality and symmetries, are often connected to the defining properties of certain mathematical objects. For instance, scattering amplitudes of gauge theories are connected to positive Grassmannian~\cite{ArkaniHamed:2012nw, Huang:2013owa} and further into the Amplituhedron~\cite{Arkani-Hamed:2013jha}, couplings of higher-dimensional operators in effective field theories, or four-point functions of a conformal field theory, are bounded by cyclic polytopes~\cite{EFT}. In each case, the mathematical object of interest has an intrinsic definition that does not make any reference to physics. In other words the physical principles become emergent from  the mathematical properties.

Recently a fascinating new connection was revealed by Galashin and Pylyavskyy~\cite{Pavel}. The observables in question are correlators of 2D planar Ising networks, which were shown to be equivalent to cells in positive Orthogonal Grassmannian. The latter was known to describe amplitudes of 3D supersymmetric Chern-Simon matter theories~\cite{Huang:2013owa, Lee:2010du}. 

In relating amplitudes to positive Grassmannian, the physical principle underlying the equivalence is factorization: start with amplitudes of the fewest particles for which the correspondence is evident, higher multiplicity amplitudes can be constructed via {\it factorization}, that is a positivity preserving operation. In this note, we identify the corresponding principle for Ising correlators: {\it amalgamation}, under which two of (adjacent) external spin sites are identified, the correlation function of the new network can be written as a non-linear function of the former. In terms of Grassmannians, the map linearizes and manifestly preserves positivity. 

The correspondence allows us to introduce recursive methods to compute correlators with large number of spin-sites by directly constructing Grassmannian. First, utilize the fact that duality moves in Grassmannian reflect dual relations between networks, we introduce iterative duality moves to construct self-similar lattices.  As the moves are equipped with maps between the couplings of two lattices, the duality map can be interpreted as an exact RG equation. Alternatively, one can employ amalgamation in the Grassmannian to construct lattices that are self repeating in one direction for arbitrary length. In this case, since the complexity of computing the move is agnostic to the underlying network, it reduces the scale of the complexity of a lattice with $N$ sites to $\sim \log N$.

\vspace{-0.4cm}

%%%%%%%%%%%%%%%%%%%%%%%%%%%%%%%%%%%%%%%%%      
\section{Mapping Ising network to cells of OG$_{\ge0}(n,2n)$}
%%%%%%%%%%%%%%%%%%%%%%%%%%%%%%%%%%%%%%%%%  
For a general Ising network, the two-point function is defined as~\cite{higherpt1} 
\begin{align}\label{sisj}
\!\!\! \langle\sigma_i\sigma_j \rangle =\frac{\sum_{\sigma_a\in\{\pm1\}}\sigma_i\sigma_j P(J_{ab})}{\sum_{\sigma_a\in\{\pm1\}} P(J_{ab})}, \,\, P(J_{ab})=\!\!\!\prod_{a,b\in\{E\}}\!\! e^{J_{ab}\sigma_a\sigma_b}\,
\end{align}
where $\sigma_i = \pm1$ represent spin sites, $E$ is the set of edges and $J_{ab} \geq 0 $ is the coupling constant connecting sites $a$ and $b$. Intuitively, since we are considering ferromagnetic couplings we expect that the correlator to be non-negative. However, as a sum with alternating signs, its positivity is not obvious. Remarkably, as proven in~\cite{Lis} that not only is  eq.(\ref{sisj}) positive, all minors of $n\times n$ unit (with $1$ in diagonal) symmetric matrix $\langle\sigma_i\sigma_j \rangle$ are positive (with definite signs)! This is referred as total positivity.

The unit symmetric matrix can be naturally embedded in a $n\times 2n$ matrix $m_{ij}$ with following map~\cite{Pavel}:
\begin{align}
&i\neq j:\quad m_{i, 2j-1}=-m_{i,2j}=Sign[i{-}j](-)^{i{+}j}\langle\sigma_i\sigma_j\rangle,\nonumber\\
&i=j: \quad m_{i, 2i-1}=m_{i,2i}=1 \,.
\end{align}
The rows of the $n\times 2n$ matrix are mutually null vectors (with alternating metric~\cite{Huang:2013owa}), termed Orthogonal Grassmannian OG$_n$. Correlators can be recovered by the inverse map, 
\eq\label{Map}
\langle \sigma_i\sigma_j\rangle=\frac{\sum_{I \in \varepsilon(\{i,j\})}\Delta_I}{\sum_{I \in \varepsilon(\{\emptyset\})}\Delta_I}\,.
\eqe 
Here $\Delta_I$ denote $n\times n$ minors of OG$_n$, and $\varepsilon(\{S\})$ represents $n$-element subsets such that for each $i$, $I\bigcap (2i{-}1,2i)$ even times if and only if $i\in S$. As the simplest example, consider the network with spin sites connected by an edge $J$, the corresponding OG$_2$ is  
\eq\label{OG24}
 \left(\begin{array}{cccc}1 & s(J) & 0 & -c(J) \\0 & c(J) & 1 & s(J)\end{array}\right)
\eqe
where 
\eq
s(J)=\frac{2}{e^{2J}{+}e^{{-}2J}}, \quad \quad c(J)=\frac{e^{2J}{-}e^{{-}2J}}{e^{2J}{+}e^{{-}2J}}.
\eqe 
With $\Delta_{12}=c(J), \Delta_{13}=1, \Delta_{14}=s(J)$, use eq.(\ref{Map}) we indeed recover the two-point function.

Note that any unit symmetric matrix can be embedded in a OG$_n$, what is special for correlators of 2D planar Ising networks is that the corresponding OG$_n$ is positive (OG$_{\geq 0, n}$)~\cite{Pavel}, i.e. all ordered $\Delta_I \geq 0$. Via eq.(\ref{Map}), the total positivity of $\langle\sigma_i\sigma_j \rangle$ can be inferred from OG$_{\geq 0, n}$.

%%%%%%%%%%%%%%%%%%%%%%%%%%%%%%%%%%%%%%%%%      
\section{The microscopic derivation of the correspondence }
%%%%%%%%%%%%%%%%%%%%%%%%%%%%%%%%%%%%%%%%%  
\begin{figure}
\begin{center}
\includegraphics[scale=0.5]{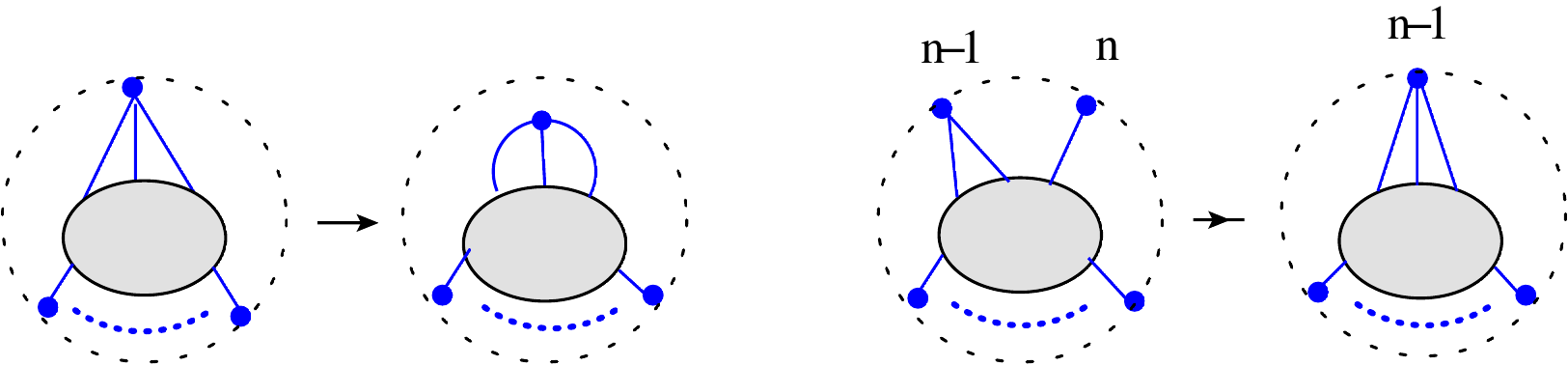}\,,
\caption{Two elementary moves that yield all Ising networks.}
\label{fig.Amalga}
\end{center}
\end{figure}
\begin{figure}
\begin{center}
\includegraphics[scale=0.2]{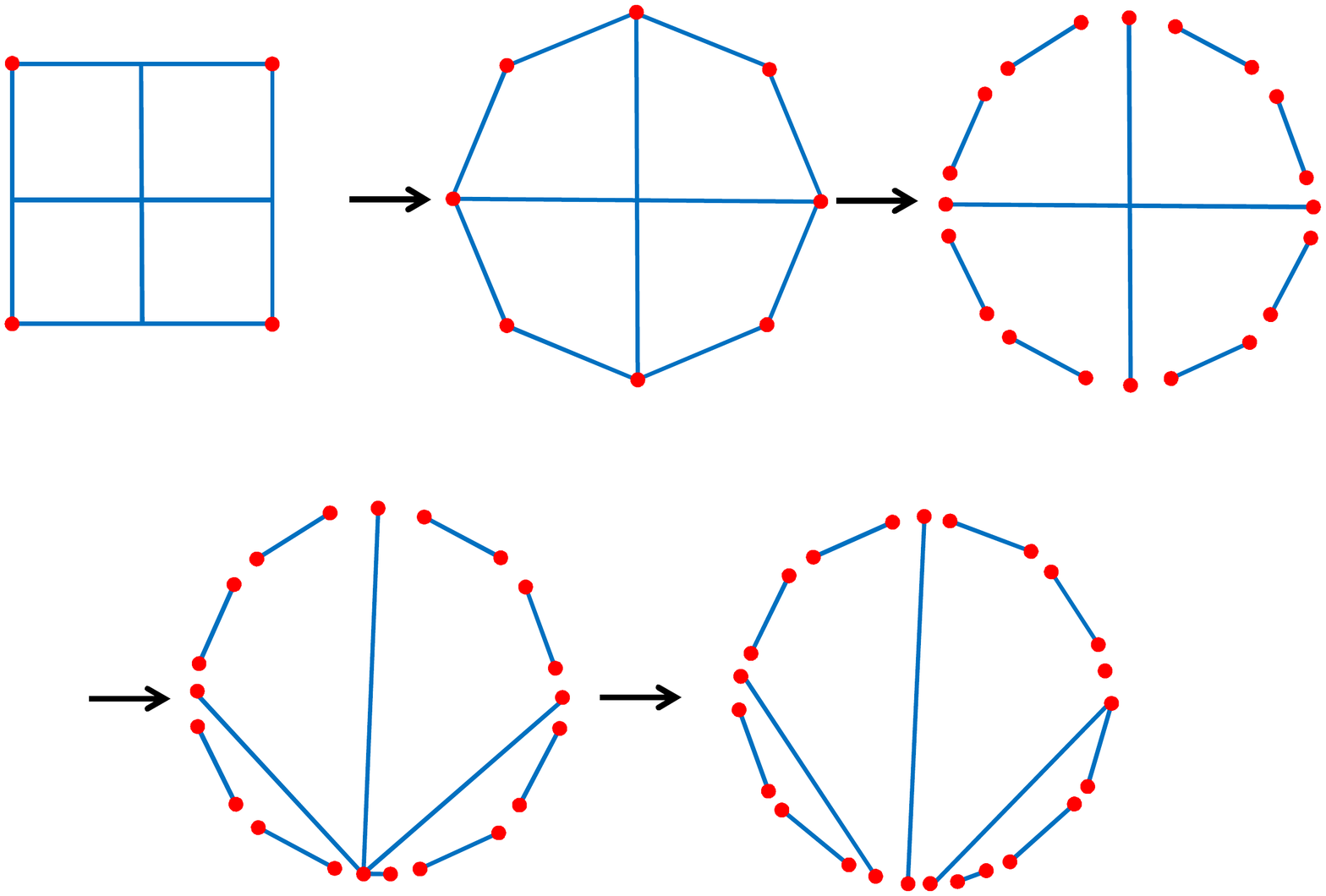}
\caption{Use the ``inverse" of the elementary moves, an Ising network can be reduced to that of only free edges.}
\label{fig.AmalgaS}
\end{center}
\end{figure}

We will establish the correspondence by showing that any Ising network can be constructed from trivial ``free-edge" networks through successive application of two elementary moves: ``pushing" external sites into the internal and the identification of two external spins, as in fig.\ref{fig.Amalga}. We will refer to the latter as {\it amalgamation}. Conversely, through the inverse, one can reduce any network to a trivial one, for example see fig.\ref{fig.AmalgaS}. 
The first move merely changes what is called external and internal, thus it does not modify correlators. For the second, there is a simple relation between correlators before and after amalgamation:
\eq\label{amalga}
\langle \sigma_i \sigma_j\rangle^{amal} =\frac{\langle \sigma_i \sigma_j\rangle+\langle \sigma_n\sigma_{n{-}1}\sigma_i \sigma_j\rangle}{1+\langle \sigma_n \sigma_{n{-}1}\rangle}\,,
\eqe
where $\langle \cdots\rangle^{amal}$ represents correlation functions of the amalgamated network. While these moves act very differently on the correlators, their images in OG$_n$ are actually identical! In both cases, we reduce the boundary sites by 1, i.e. OG$_n$ to OG$_{n{-}1}$. For the first case, we get the same correlator by embedding either in OG$_n$ or OG$_{n{-}1}$. For instance, consider from OG$_3$ to OG$_2$, the statement that the same $\langle \sigma_1 \sigma_2\rangle$ can be obtained from either embedding leads to
\eq \label{eq:2pt2}
\langle \sigma_1 \sigma_2\rangle=\frac{\Delta^{(2)}_{12}}{\Delta^{(2)}_{13}{+}\Delta^{(2)}_{14}}=\frac{\Delta^{(3)}_{125}{+}\Delta^{(3)}_{126}}{\Delta^{(3)}_{135}{+}\Delta^{(3)}_{136}{+}\Delta^{(3)}_{145}{+}\Delta^{(3)}_{146}}\,,
\eqe  
where $\Delta_I^{(n)}$ are minors of OG$_n$. The equality implies that minors of two Grassmannians are related via:
\eq\label{amalgaOG}
\Delta^{(n{-}1)}_{\{I\}}=\Delta^{(n)}_{\{I a\}}+\Delta^{(n)}_{\{I b\}}\,,
\eqe 
where the columns $\{a,b\}$ ($\{5,6\}$ here) correspond to that of the spin ``pushed" into the internal. Next we consider amalgamation of identifying sites $2$ and $3$ to reduce OG$_3$ to OG$_2$. The $3\times6$ matrix of OG$_3$ are labelled as %$m_{ij}$
%\eq
%\left(\begin{array}{cccccc}\uparrow & \uparrow & \uparrow & \uparrow & \uparrow & \uparrow \\ c_1 & c_2 & c_3 & c_a & c_b &c_4 \\ \downarrow & \downarrow & \downarrow & \downarrow & \downarrow & \downarrow\end{array}\right)\,.
%\eqe
$$\includegraphics[scale=0.2]{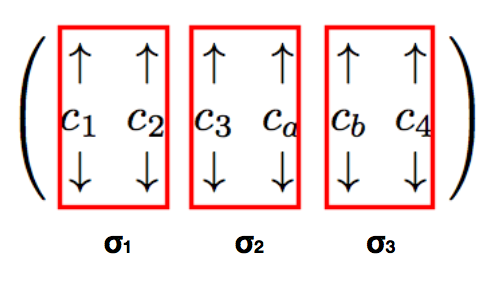}$$
Now the extra columns $\{a,b\}$ are two adjacent columns of spin sites $\sigma_2$ and $\sigma_3$. Use the relation eq.(\ref{amalga}),  eq.(\ref{Map}) again implies eq.(\ref{amalgaOG})! Thus fundamental moves become identical when embedded in OG$_n$, up to the positions of the removed columns. This is reminiscent to conformal symmetry~\cite{Elvang}: translation and conformal boosts are drastically different and non-linear in general, but they are unified and linearized in twistor space.    

Importantly, minors of new network are a \textit{positive} sum of those of the old one: fundamental moves preserve the positivity! Thus the positivity of general networks boils down to the property of ``free-edge" networks, whose Grassmannian is simply the embedding of multiple OG$_{\ge0,2}$s. For ordered planar networks, it is just a block embedding,
\eqqa
\includegraphics[scale=0.35]{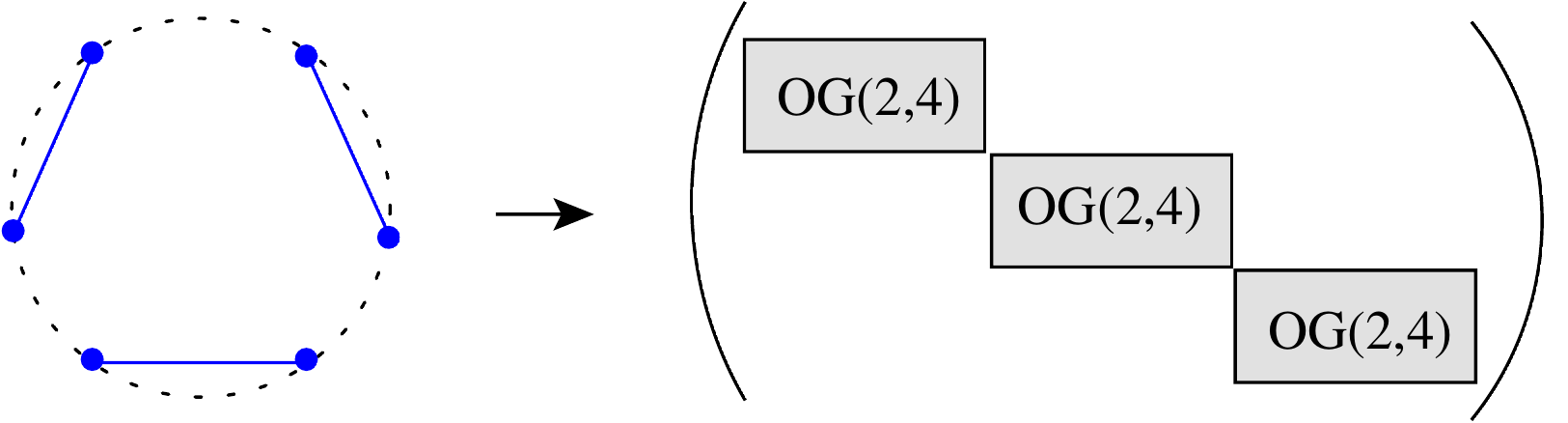}\,
\eqqae 
Non-vanishing minors require exactly two columns from each OG$_{\ge0,2}$, they are positive products of minors of OG$_{\ge0,2}$, which are manifestly positive. The embeddings of general \textit{planar} ``free-edge" networks are simply even permutations, of the columns, away from block embeddings. Thus they all live in OG$_{\ge0,n}$, and the fundamental moves lead to the correspondence for general networks.

%%%%%%%%%%%%%%%%%%%%%%%%%%%%%%%%%%%%%%%%%      
\section{The structure of OG$_{\ge0,n}$ and Equivalence moves}
%%%%%%%%%%%%%%%%%%%%%%%%%%%%%%%%%%%%%%%%%  

The space of OG$_{\ge0,n}$ consists of {\it cells}, each can be represented by an on-shell diagram constructed by quartic vertices. Two diagrams are equivalent if they are related by {\it equivalence moves} \cite{ArkaniHamed:2012nw} via a change of variables, which consists of {\it bubble reductions} and {\it triangle move} for OG$_{\ge0,n}$ \cite{Huang:2013owa, Huang:2014xza}. The images of these moves in Ising network were actually recognized long ago~\cite{old-refs.}. The change of variables of on-shell diagrams then corresponds to a map between couplings of Ising networks that are dual.

There are two kinds of Ising networks whose corresponding on-shell diagrams contain a bubble. First: 
\eqqa \label{fig:bubble_reduction2}
\begin{picture}(190, 81)
\includegraphics[scale=0.15]{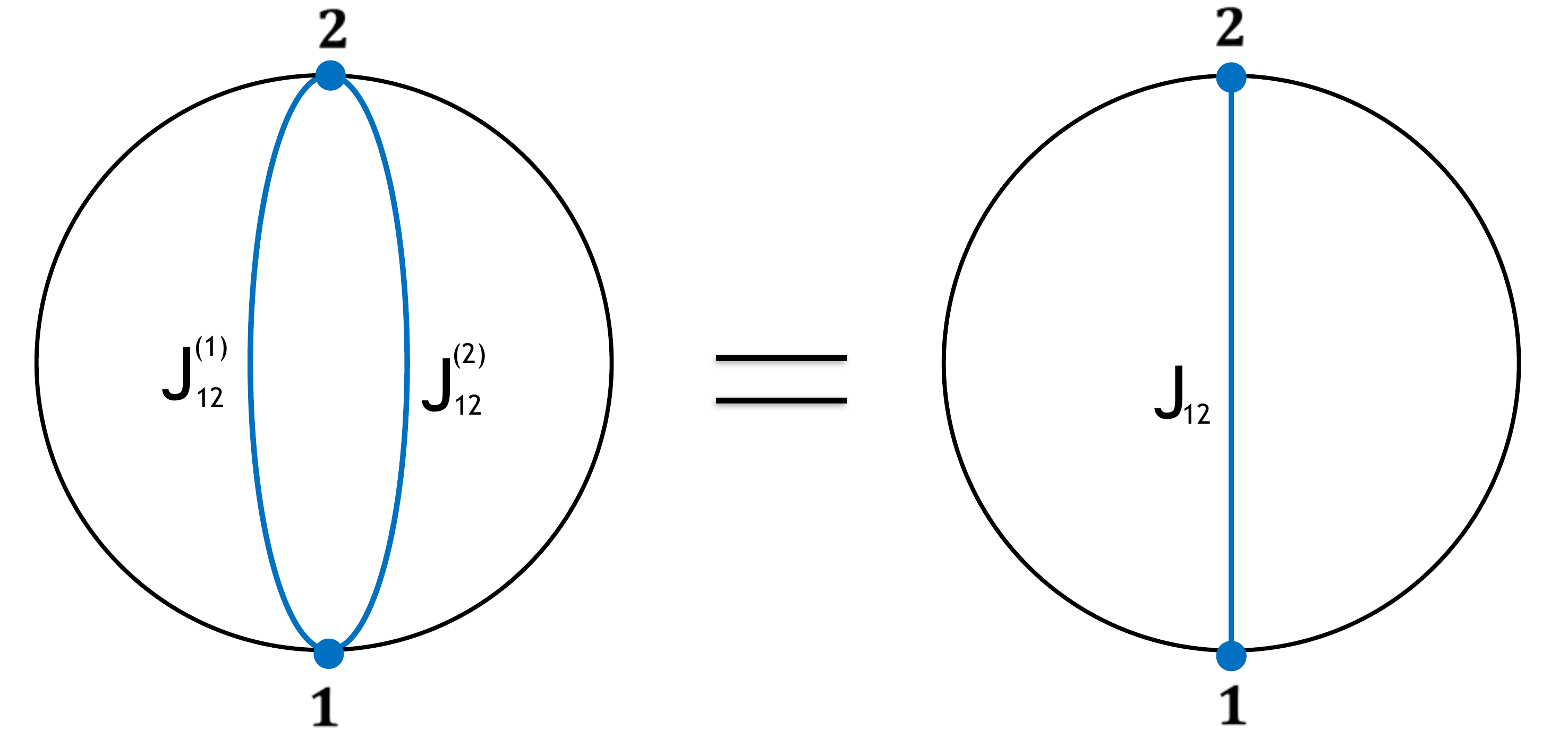} 
\end{picture}
\eqqae
As evident from the graph we simply have $J_{12}= J^{(1)}_{12}+J^{(2)}_{12}$. Another kind of graph is given by, 
\eqqa \label{fig:bubble_reduction1}
\begin{picture}(190, 79)
\includegraphics[scale=0.16]{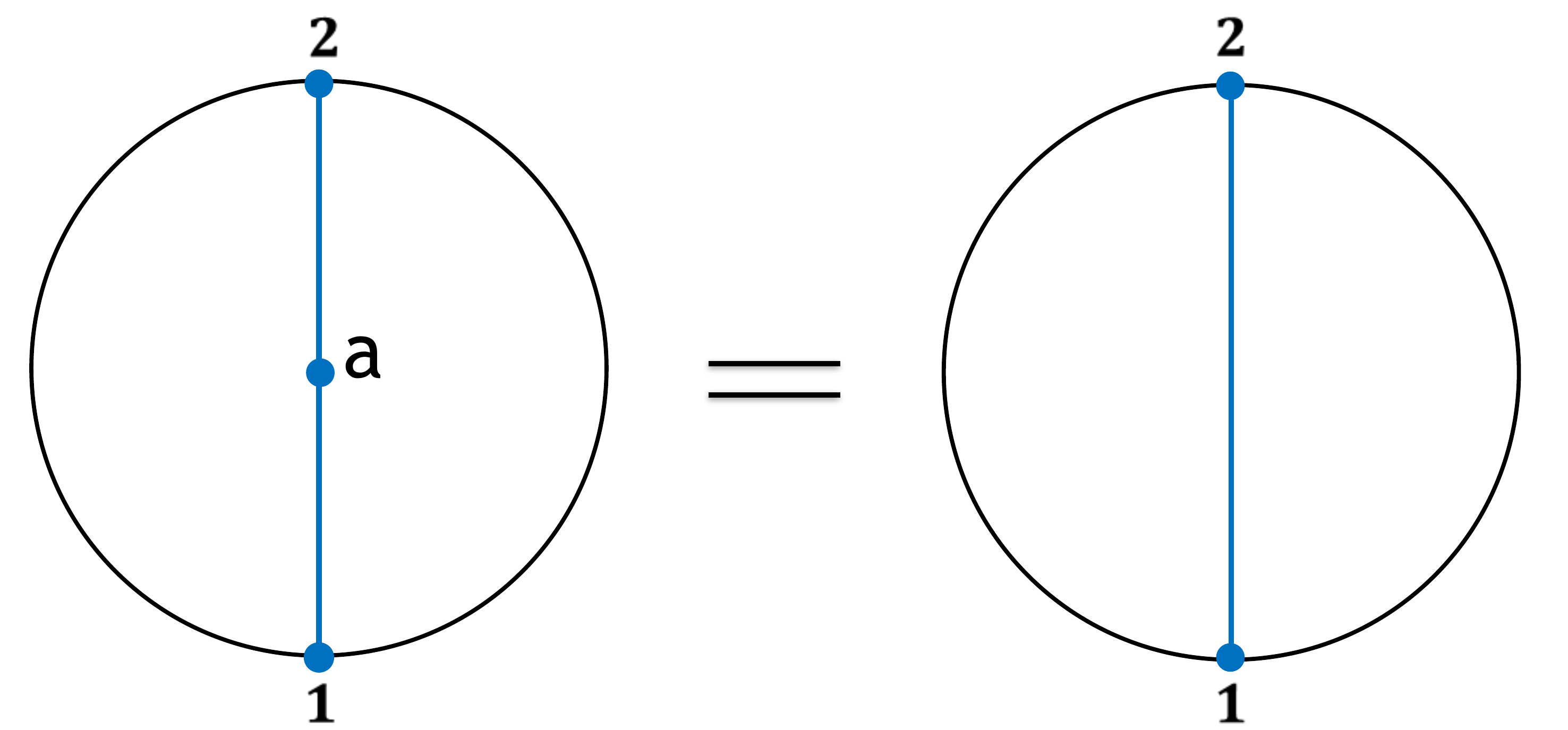} 
\end{picture}
\eqqae
In this case, the reduction is to remove the isolated spin $a$ and to define an effective coupling via~\cite{Huang:2014xza} 
\begin{align} \label{eq:ABJMbubble1}
c(J_{12}) &= {c(J_{1a}) c(J_{2a}) \over 1+ s(J_{1a}) s(J_{2a})} , \cr
 s(J_{12}) &= {s(J_{1a}) + s(J_{2a}) \over 1+ + s(J_{1a}) s(J_{2a})} \, .
\end{align}
The {\it triangle move} relates two triangle on-shell diagrams: 
\eqqa
\begin{picture}(180, 75)
\includegraphics[scale=0.19]{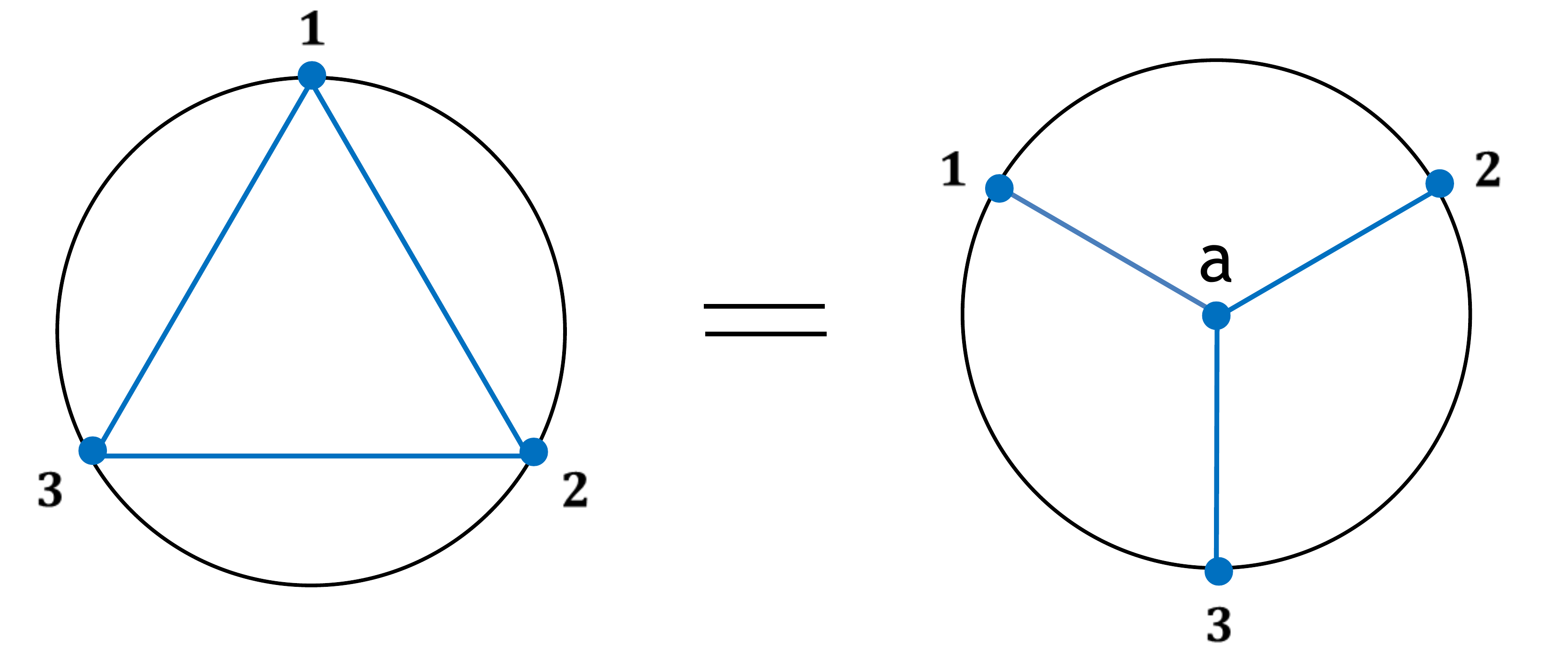} 
\end{picture}
\eqqae
The duality transformation is given by \cite{Huang:2014xza} 
\begin{align} \label{eq:ABJM_triangle}
s(J_{ia}) &= { s(J_{i i+1}) c(J_{i+1\, i+2}) s(J_{i\,i+2})  \over c(J_{i+1\, i+2})+ c(J_{i i+1}) c(J_{i\, i+2})} \, ,  \cr
 c(J_{i i+1}) &= {c(J_{ia}) c(J_{i+1\, a}) s(J_{i+2\,a}) \over s(J_{i+2\,a}) + s(J_{ia}) s(J_{i+1\,a})}, 
\end{align}
for $i=1,2,3$ with $i+3:=i$ is understood.

\vspace{-0.4cm}
%%%%%%%%%%%%%%%%%%%%%%%%%%%%%%%%%%%%%%%%%      
\section{Recursion relations through the Grassmannian}
%%%%%%%%%%%%%%%%%%%%%%%%%%%%%%%%%%%%%%%%%  
Since the information of  correlation functions of Ising networks are completely captured by OG$_{\ge0,n}$, one obtains all the correlators by constructing the Grassmannian. This leads to new methods to compute correlators. 
\vspace{-0.4cm}
    \subsection{Recursions via duality transformations}
When the equivalence move are applied to a network with self-similar structure, the map for effective couplings become recursion relations of exact RG equation type. To illustrate the idea we consider some examples. Begin with the Sierpinski triangle:
\eqqa \label{fig:Sierpinskitriangle}
\begin{picture}(450, 65)
\includegraphics[scale=0.17]{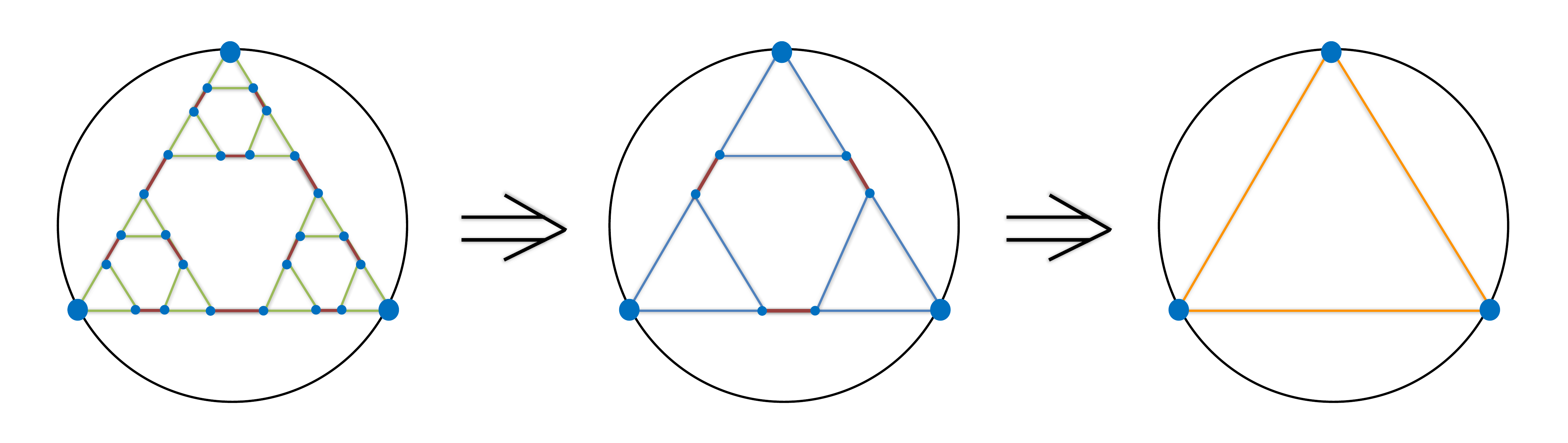} 
\end{picture}  \nonumber
\eqqae
It is a system with two kinds of couplings, marked with different colors: the one connecting between triangles (called $J_1$ and marked with brown), whereas the rest is $J_2$.  Use the duality transformations, we obtain the recursion relation for $J_2$, 
\begin{align}  \label{eq:sstriangle2}
c(J_2') ={ c(J_1)\, c(J_2)^2 \over c(J_2)^2 +2(1-c(J_2)) (1+s(J_1) - {1\over 2}c(J_1)\, c(J_2)) } \, .
\end{align}
In the limit $J_1 \rightarrow \infty$, namely we shrink all the brown edges, the above relation reduces to
\begin{align} 
c(J'_2) ={ c(J_2)^2 \over 2c(J_2)^2 -3 c(J_2)+2} \, ,
\end{align}
which agrees with the result in~\cite{recursion_triangle, triangle}. Another example we consider here is: 
\begin{align} \label{fig:Triangle_triangle}
\begin{picture}(440, 68)
\includegraphics[scale=0.17]{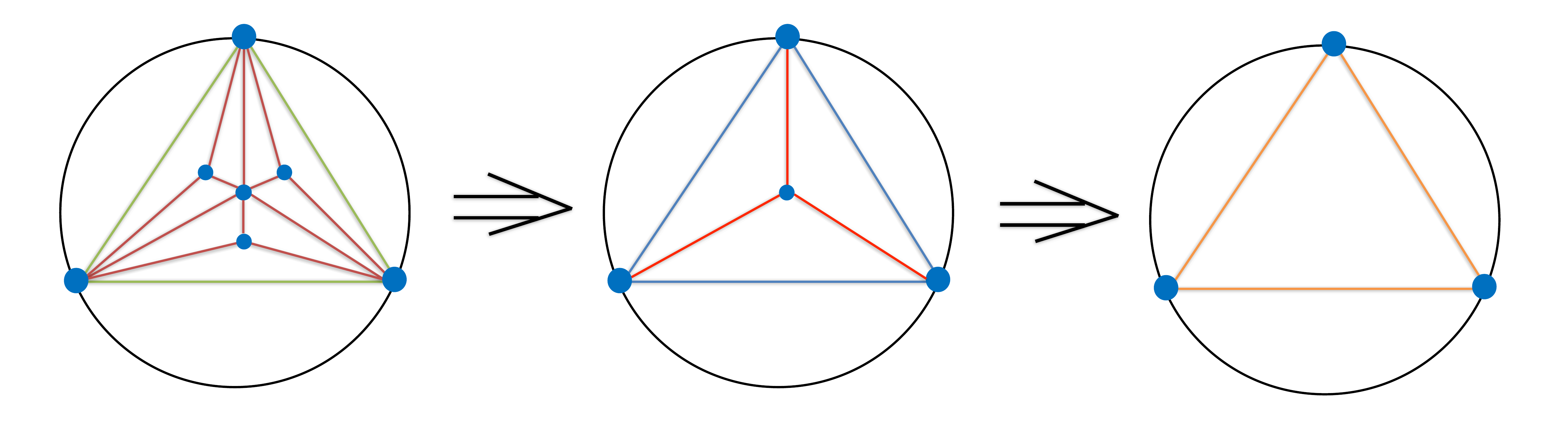}
\end{picture}  \nonumber
\end{align}
It is constructed by starting with a triangle, then triangulates it into $3$ smaller ones, and continues with the same procedure. As shown again by colors, the outer edges have same coupling $J$, whereas all the internal ones being $J_I$. The structure remains under the recursions. Focus on the last step of the recursion, we have, 
\begin{align} 
c(J') ={1-s(J_I)+c(J) \over 1+ c(J) (1-s(J_I))} \, .% \label{eq:bubble_reduction3}
\end{align} 
We comment that the fixed points to the recursions discussed here are all simply $J=0$ or $\infty$, thus the systems do not exhibit finite temperature phase transitions. 
\vspace{-0.4cm}
    \subsection{Recursions through amalgamation}
\begin{figure}
\begin{center}
\includegraphics[scale=0.25]{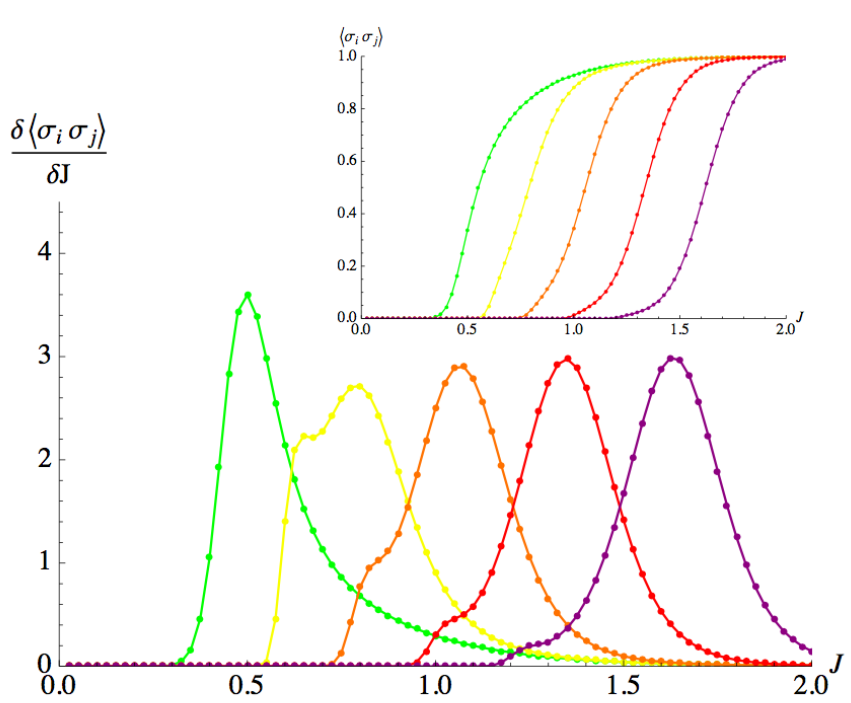}
\caption{The variation of the two-point function with respect to the coupling $J$ for the network in eq.(\ref{Recurs}), with iterations 1, 5, 10, 15, 20, and 25, displayed successively to the right. The inset is the plot of the two point function itself. }
\label{Plots}
\end{center}
\end{figure}

The amalgamation is agnostic to the underlying Grassmannian. Thus when apply recursively the same construction to build a large network, the complexity is constant at each iteration. For an OG$_{\ge0,2n}$, corresponding to any network with $2n$ boundary sites and $N_0$ total sites,  amalgamating with itself along $n$ edges leads to
\eqqa\label{Recurs}
\includegraphics[scale=0.36]{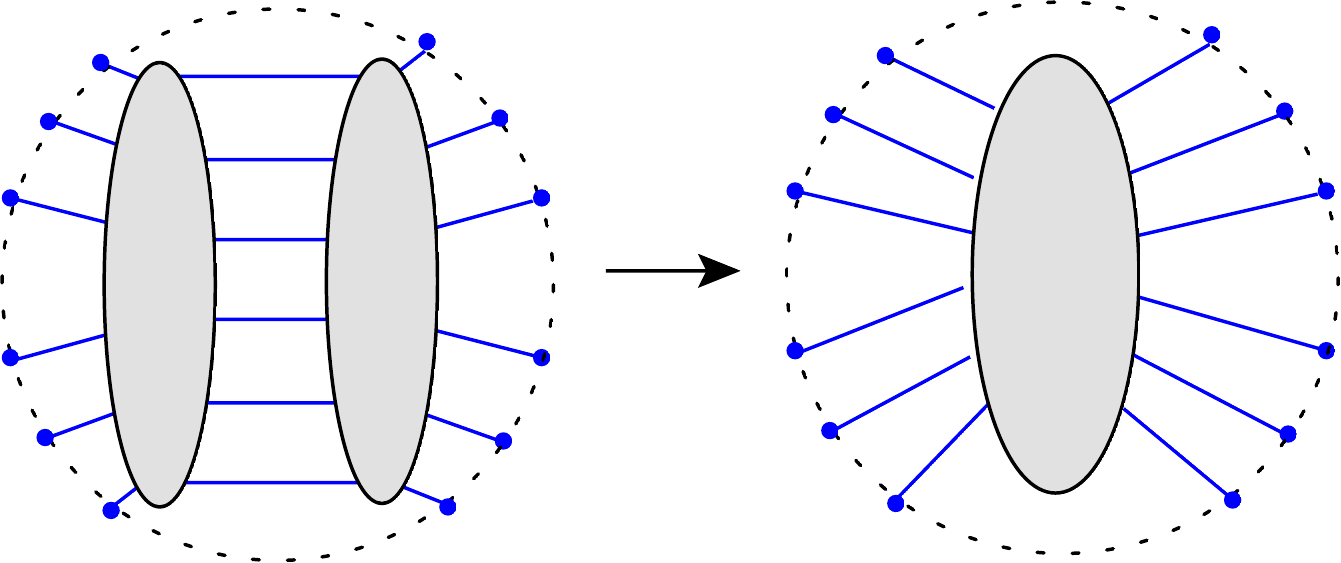} 
\eqqae
The result is a new OG$_{\ge0,2n}$ with $2N_0-n$ total sites. We note that if the final lattice contains $N$ sites, the computation complexity scales as $\log N/N_0$ using the iteration method, in contrast to the linear growth for the conventional approach. As an illustration of its power, consider the iteration of the following network:  
\eqqa
\begin{picture}(230, 150)
\includegraphics[scale=0.2]{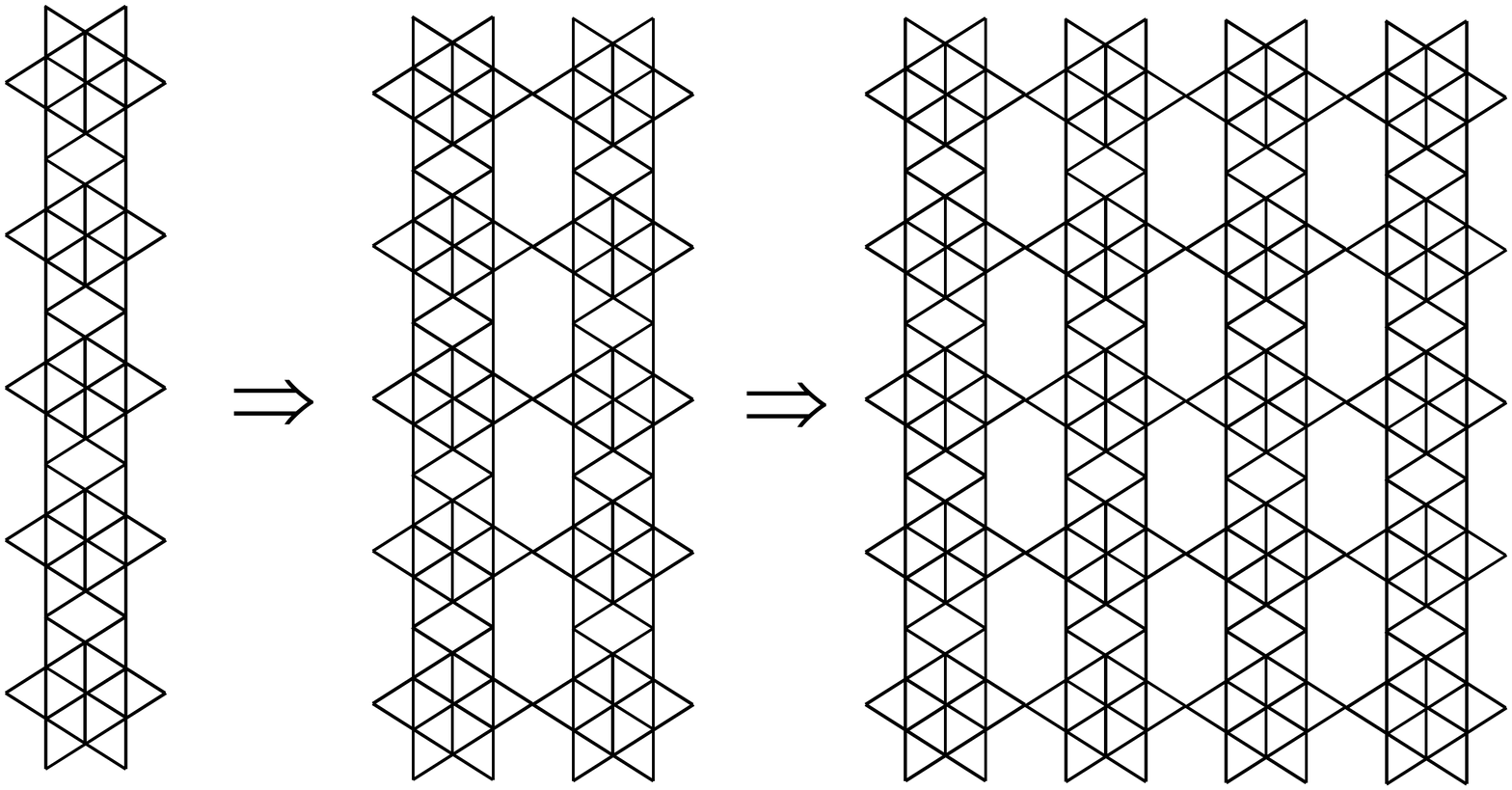}  \nonumber
\end{picture} 
\eqqae
where $N_0=57$. Applying straightforwardly the recursion up to $25$ iterations results  in over $1.7\times 10^{9}$ spin sites. The correlator with respect to coupling $J$, as well as its tangent slope are plotted in fig.\ref{Plots}. We see that the tangent slope stabilizes under iteration, showing no first-order phase transition. 
\subsection{Phase transitions}
The above two recursive constructions of correlators are applicable to lattices which have ``finite ramification": those can be partitioned by removing finite vertices. For the duality based recursions, by construction the final result is dual to a simple finite lattice, while for the amalgamation recursion the result is simply a sum of finite lattices, so one should not observe any phase transitions. This shows the lack of phase transitions for lattices with finite ramification numbers~\cite{triangle}. 
%%%%%%%%%%%%%%%%%%%%%%%%%%%%%%%%%%%%%%%%%      
%\subsection{The reduced networks}
%%%%%%%%%%%%%%%%%%%%%%%%%%%%%%%%%%%%%%%%%
\vspace{-0.2cm}

%%%%%%%%%%%%%%%%%%%%%%%%%%%%%%%%%%%%%%%%%      
\section{Conclusions and Outlook}
%%%%%%%%%%%%%%%%%%%%%%%%%%%%%%%%%%%%%%%%%
We explore the correspondence between OG$_{\ge0,n}$ and 2D planar Ising networks, which is established via fundamental moves and positivity of the simplest free-edge networks. The correspondence leads to duality transformations that relate networks in the same cell of Grassmannian, and the amalgamation construction for general networks. Duality transformations and amalgamation are actually applicable beyond 2D planar networks, while positivity of free-edge networks holds even with external magnetic fields. It is of interest to explore to what extent the results in this paper can be applied.

%%%%%%% Appendix %%%%%%%

\vspace{-0.4cm}

%%%%%%%%%%%%%%%%%%%%%%%%%%%%%%%%%%%%%%%%%%%%%%%%%%%%%
\section{Acknowledgements}
%%%%%%%%%%%%%%%%%%%%%%%%%%%%%%%%%%%%%%%%%%%%%%%%%%%%
We thank P. Galashin and P. Pylyavskyy for bringing to our attention their fascinating work. We also like to thank Nima Arkani-Hamed for very enlightening comments. C.W. is supported by a Royal Society University Research Fellowship no. UF160350. C-k Kuo and Y-t Huang is supported by  MoST Grant No. 106-2628-M-002-012-MY3.

%{99}


\begin{thebibliography}{99}
\bibitem{higherpt1}  
Higher-point functions can be written as products of two-point functions, and thus the correspondence generalizes easily~\cite{Pavel}.
\end{thebibliography}
\end{document}